\documentclass[10pt,conference]{IEEEtran}
\IEEEoverridecommandlockouts
\usepackage{cite}

\usepackage{hyperref}
\usepackage{url}
\hypersetup{
    colorlinks=true,
    linkcolor=black,
    urlcolor=blue,
    citecolor=black,
}
\usepackage{amsfonts}
\usepackage{paralist}
\usepackage[utf8x]{inputenc}
\usepackage{amsmath}
\usepackage{pifont}
\usepackage{graphicx}
\usepackage{xspace}
\usepackage[ruled,linesnumbered]{algorithm2e}
\usepackage{multirow}
\usepackage{tabularx}
\usepackage{colortbl}
\usepackage[caption=false,font=footnotesize]{subfig}
\usepackage{amsthm}

\newtheorem*{definition}{Definition}

\definecolor{LightCyan}{rgb}{0.88,1,1}
\definecolor{Gray}{gray}{0.9}

\newcommand{\OSS}{FOSS\xspace}
\newcommand{\code}[1]{\textsf{\small#1}}
\long\def\longcaption#1#2{\centering\begin{minipage}{#1}\footnotesize\vspace{0.1\baselineskip}\noindent\emph{#2}\vspace{0.1\baselineskip}\end{minipage}}

\usepackage{wrapfig}

\newenvironment{bio}[1]
{\par
 \bigskip
 \begin{wrapfigure}{l}[0pt]{1in}
 \vspace{-15pt}
 \includegraphics[height=1in,clip,keepaspectratio]{#1}
 \vspace{-25pt}
 \end{wrapfigure}
 \footnotesize \noindent}
{\par\bigskip}

\begin{document}

\onecolumn

\begin{figure}
    \centering
    \includegraphics[width=.3\textwidth]{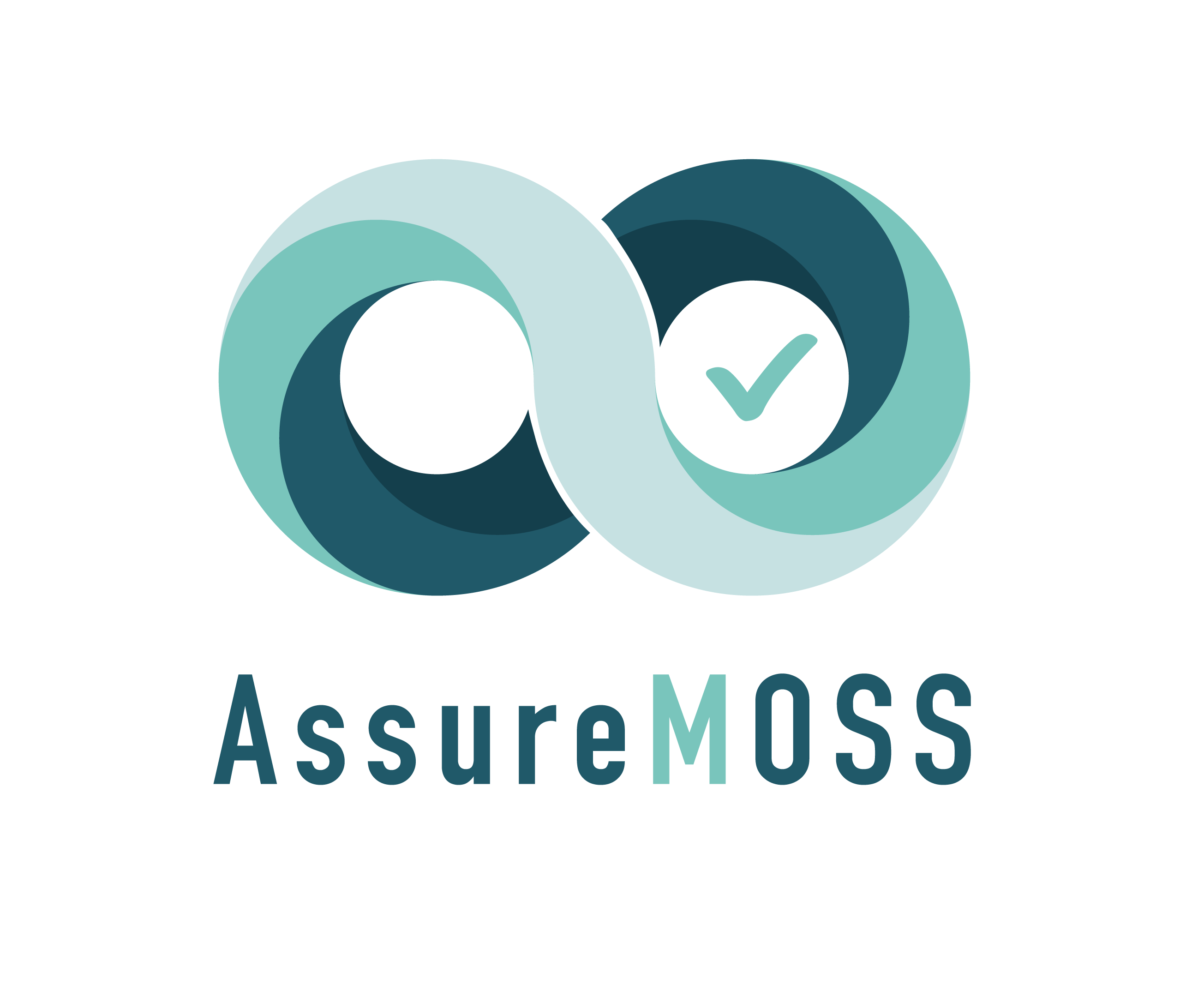}
\end{figure}

\vspace{2\baselineskip}

\begin{figure}[h!]
    \centering
    \includegraphics[width=\textwidth]{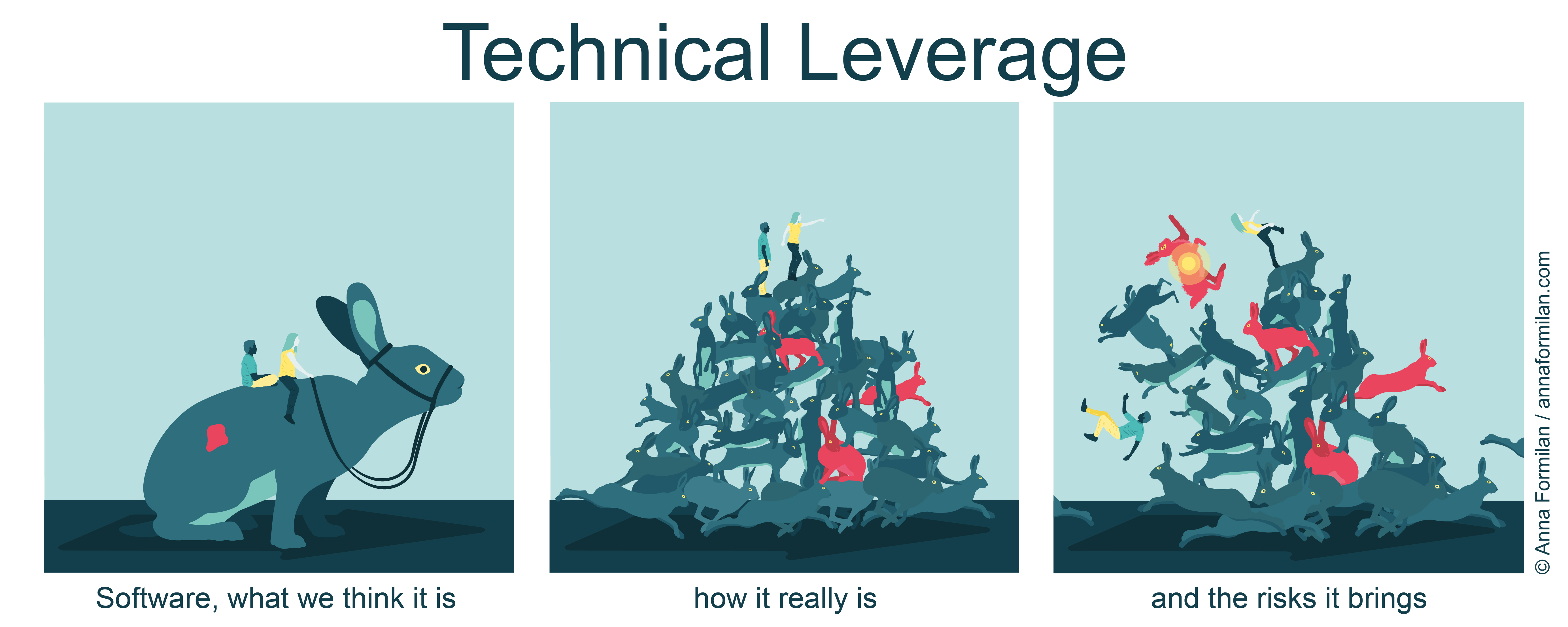}
\end{figure}

\vspace{2\baselineskip}

\begin{center}
{\huge \textbf{Technical Leverage in a Software Ecosystem: Development 
Opportunities and Security Risks}}
\end{center}

\vspace{\baselineskip}

{\large
Authors:
\begin{itemize}
	\item[] \textbf{Fabio Massacci}, University of Trento (IT), Vrije Universiteit Amsterdam (NL)    
    \item[] \textbf{Ivan Pashchenko}, University of Trento (IT)
\end{itemize}
}

\vfill

\begin{wrapfigure}{l}{2.5cm}
\vspace{-\baselineskip}
\includegraphics[width=2.5cm]{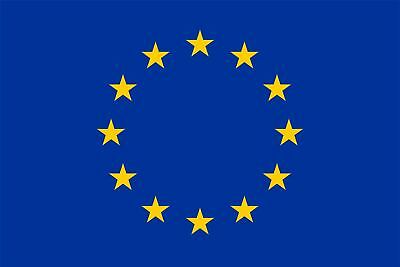}
\end{wrapfigure}

This paper was written within the H2020 AssureMOSS project that received funding 
from the European Union's Horizon 2020 research and innovation programme under 
grant agreement No 952647. This paper reflects only the author's view and the 
Commission is not responsible for any use that may be made of the information 
contained therein.

\clearpage
\twocolumn
\begin{bio}{graph/Logo_colori}
\textbf{Assurance and certification in secure Multi-party Open Software and 
Services (AssureMOSS)} No single company does master its own national, in-house 
software. Software is mostly assembled from “the internet” and more than half 
come from Open Source Software repositories (some in Europe, most elsewhere). 
Security \& privacy assurance, verification and certification techniques 
designed for large, slow and controlled updates, must now cope with small, 
continuous changes in weeks, happening in sub-components and decided by third 
party developers one did not even know they existed. AssureMOSS proposes to 
switch from process-based to artefact-based security evaluation by supporting 
all phases of the continuous software lifecycle (Design, Develop, Deploy, 
Evaluate and back) and their artefacts (Models, Source code, Container images, 
Services). The key idea is to support mechanisms for lightweigth and scalable 
screenings applicable automatically to the entire population of software 
components by Machine intelligent identification of security issues, Sound 
analysis and verification of changes, Business insight by risk analysis and 
security evaluation. This approach supports fast-paced development of better 
software by a new notion: continuous (re)certification. The project will 
generate also benchmark datasets with thousands of vulnerabilities. AssureMOSS: 
\textbf{Open Source Software: Designed Everywhere, Secured in Europe}. More 
information at \textbf{\url{https://assuremoss.eu}}.
\end{bio}

\begin{bio}{photos/massacci}
\textbf{Fabio Massacci} (Phd 1997) is a professor at the University of 
Trento, Italy, and Vrije Universiteit Amsterdam, The Netherlands. 
He received the Ten Years Most Influential Paper award by the IEEE Requirements Engineering Conference in 2015. 
He is the the European Coordinator of the AssureMOSS project. Contact him at \emph{fabio.massacci@ieee.org}.
\end{bio}

\begin{bio}{photos/pashchenko}
\textbf{Ivan Pashchenko} (PhD 2019) is a Research Assistant Professor at the 
University of Trento, Italy. He was awarded a silver medal at the ACM/Microsoft 
Student Graduate Research Competition at ESEC/FSE. He is UniTrento main contact 
in \textit{``Continuous analysis and correction of secure code''} work package 
for the AssureMOSS project. Contact him at \emph{ivan.pashchenko@unitn.it}.
\end{bio}

How to cite this paper:
\begin{itemize}
    \item Massacci, F. and Pashchenko, I. Technical Leverage in a Software 
Ecosystem: Development Opportunities and Security Risks. \emph{Proceedings of 
the International Conference on Software Engineering (ICSE 2021)}. IEEE Press.
\end{itemize}

License:
\begin{itemize}
\item This article is made available with a perpetual, non-exclusive, non-commercial license to distribute.
\item The graphical abstract is an artwork by Anna Formilan.
\end{itemize}

\clearpage

\title{Technical Leverage in a Software Ecosystem: Development 
Opportunities and Security Risks}

\author{\IEEEauthorblockN{Fabio Massacci}
\IEEEauthorblockA{University of Trento (IT), Vrije Universiteit Amsterdam (NL) \\
fabio.massacci@ieee.org}
\and
\IEEEauthorblockN{Ivan Pashchenko}
\IEEEauthorblockA{University of Trento (IT) \\
ivan.pashchenko@unitn.it}
}

\maketitle
\thispagestyle{plain}
\pagestyle{plain}

\begin{abstract}
In finance, \emph{leverage} is the ratio between assets 
borrowed from others and one's own assets. A matching situation is present in 
software: by using free open-source software (\OSS) libraries a developer 
leverages on other people's code to multiply the offered 
functionalities with a much smaller own codebase. In finance as in software, 
leverage magnifies profits when returns from borrowing exceed costs 
of integration, but it may also magnify losses, in particular in the presence of 
security vulnerabilities.
We aim to understand the level of \emph{technical leverage} in the \OSS ecosystem and 
whether it can be a potential source of security vulnerabilities.
Also, we introduce two metrics \textit{change distance} 
and \textit{change direction} to capture the amount and the evolution of the dependency 
on third-party libraries.

The application of the proposed metrics on 8494 distinct library versions
from the \OSS Maven-based Java libraries shows 
that small and medium libraries (less than 100KLoC) have disproportionately more 
leverage on \OSS dependencies in comparison to large libraries.
We show that leverage pays off as leveraged libraries only add a 4\% delay in 
the time interval between library releases while providing four times more code than their own.  However,  
libraries with such leverage (i.e., 75\% of libraries in our sample) also have 
1.6 higher odds of being vulnerable in comparison to the libraries with 
lower leverage. 

We provide an online demo for computing the proposed metrics for real-world 
software libraries available under the following URL: 
\url{https://techleverage.eu/}.
\end{abstract}


\section{Introduction}
\label{sec:ecosystems:intro}

Finance and software have always found
interesting correspondences between concepts \cite{techdebtmetaphor2012}.
For example, the notion of \textit{technical debt} captures the short-term developers' action (actually inaction to fix) 
that may lead to a later cost of maintenance~\cite{avgeriou2016managing}. Most academic studies (see a review 
in~\cite{ampatzoglou2015financial}) 
only consider poorly written \emph{own} code of a project as a 
source of a technical debt for a software project. At the Dagstuhl Seminar 
16162, researchers reviewed and consolidated the view on technical 
debt~\cite{avgeriou2016managing} to clearly limit its scope to the ``internal system qualities'' of a software project.

Nowadays developers often import functionality 
from third-party free open-source software (\OSS) libraries by including them into their projects as 
dependencies~\cite{bird2015art, grinter1996understanding, dilawer2011practical}. Such software engineering practice allows developers to 
use \OSS libraries as building blocks, and therefore, reduce development cost 
and time~\cite{mohagheghi2007quality}. Even for proprietary software, the fraction of homegrown code decreased to 
5\%\cite{midlife}. Industry reports show that third party code inherited through 
dependencies is four times larger than the size of the own code base as an 
industry average~\cite{blackduck}. It can be up to four orders of magnitude in our \OSS sample.

To capture this phenomenon, we introduce the 
notion of \textbf{technical leverage} to assess 
the dependence on third-party functionalities. 
Similarly to the financial ratio between debt (other 
people's money) and equity (one's own money),
leverage is the ratio between third party code and 
one's own code. 

Additionally, we introduce the \emph{change distance} and 
\emph{change direction} metrics to measure the qualitative changes of a 
library between two consecutive versions. These metrics capture the polar coordinates of the changes in the 
plane described by the sizes of own code and third party code. An angle 
of 90 degrees means that in the new version the developer is improving its own `capital' (i.e.,\ 
code) while keeping the same third party code on which
is old version is leveraged.

Since developers have different strategies~\cite{pashchenko2020qualitative}, we are 
interested to check whether the proposed metrics characterizes 
in some way the \OSS\ ecosystem.
For example, developers of a small library may want to increase 
functionality of their library as fast as possible, while developers of a mature and large
library (i.e., with more than 100KLoC) might likely focus on fixing bugs and vulnerabilities and refining functionalities. 
\begin{compactitem}
\item[\bf RQ1:] \emph{Is there a difference in leverage, distance and direction of changes between small and large libraries?}
\end{compactitem}

The next two research questions focus on the trade-off between risk and opportunity that 
leverage may bring. For example, a large leverage means that several libraries are used and 
they might require integration and update costs. Indeed, developers often decide not to 
update the third party libraries they are using~\cite{cox2015icse, pashchenko2018vulnerable} 
due to the possibility of introducing 
incompatible, breaking changes~\cite{pashchenko2020qualitative}. Thus leverage might 
significantly delay the releases of one's own code. If that happened, the opportunity of 
leverage would turn out to be not a real opportunity but only an illusion since
the time interval between library releases is
linked to profitability~\cite{august2013influence} and costs~\cite{huijgens2017effort}.
\begin{compactitem}
\item[\bf RQ2 - Opportunity:] \emph{do leverage,  distance,  and direction of changes impact the
time interval between library releases?}
\end{compactitem}

On the risk side, using many
libraries increases the attack surface, and third-party libraries are known to 
introduce functionality bugs and security vulnerabilities into the  
projects that use them~\cite{kula2017ese,pashchenko2018vulnerable}. In some 
cases, dependent projects keep using outdated components for a decade or 
more~\cite{dashevskyi2018effort} thus
increasing also the window of possible exploitations.
\begin{compactitem}
\item[\bf RQ3 - Risk:] \emph{does leverage, distance and direction of changes impact the risk of including vulnerabilities?}
\end{compactitem}  

To answer the RQs above, we applied the 
proposed metrics to more than 10K distinct library instances
used in the  \OSS\ Java Maven-based ecosystem distinguishing
between large libraries (over 100KLoC) and small-medium libraries.

\noindent\textbf{Summary of Findings} Our analysis suggests that small-medium libraries (less than 100KLoC) 
have a high leverage on third-party \OSS 
dependencies and their developers prefer to adopt new dependencies (change direction at 0 degrees). Large libraries have far smaller 
leverage and their developers mostly increase their own code.
The proposed metrics also partly explain the time interval between library releases but even a large leverage 
(more than 4 times one's own code,  which is present
in 75\% of the small and medium libraries in our sample) 
yield a very minimal change to such interval (less than 4\%).  Leverage is 
thus a concrete \emph{opportunity for the developers} of a library. 
Yet, a large leverage also increases 
the odds ratio of shipping code with vulnerabilities by 60\%. So leverage 
is an equally concrete \emph{risk for the users} of the library. 

Such findings bring new challenges for empirical software engineering (your `software' and hence its quality depends way more from your choices of third party libraries than your coding) and software security economics
(code users bear the risk while code developers reap the benefits).

\section{Terminology}
\label{sec:terminology}

We rely on the terminology established among practitioners (e.g., the users of Apache Maven) and 
consolidated in~\cite{pashchenko2018vulnerable}:
\begin{itemize}
\item A \textit{library} is a separately distributed software component, typically consisting of a 
logically grouped set of classes (objects) or methods (functions). To avoid ambiguity, we refer
to a specific version of a library as a \textit{library instance}.
\item A \textit{dependency} is a library instance whose functionalities are used by another library instance (\textit{dependent} instance).
\item A \textit{direct} dependency is \textit{directly} invoked from the dependent library instance.
\item A \textit{dependency tree} is a representation of a library instance where 
each node is a library instance and edges connect dependent library instances to their direct dependencies.
\item A \emph{transitive dependency} of a library instance at a root of a dependency tree is connected to the root library through a path with 
more than one edge.
\item A \textit{project} is a set of libraries developed and/or maintained together by a group of developers. 
Dependencies belonging to the same project of the dependent library instance are \textit{own 
dependencies}, while library instances maintained by other projects are \textit{third-party 
dependencies}.
\end{itemize}

Additionally, for each library instance in our sample we identify the following dimensions that characterize a library:
\begin{itemize}
    \item \textit{Own code size} ($\ell_{own}$) as the number of lines of own code in the files of a 
library.
    \item \textit{Dependency code size} ($\ell_{dep}$) as the sum
    of the lines of code of \emph{third-party} direct dependencies ($\ell_{dir}$)  and 
transitive dependencies ($\ell_{trans}$)  of a library
    \item \textit{Total code size} ($\ell_{total}$) as the sum of the two.
\end{itemize}
The qualifier `third party' is important as noted by Pashchenko et al. 
\cite{pashchenko2018vulnerable}: for convenience, developers might have decided 
to structure their own code in separate libraries. They might be mistakenly 
counted as other people's code while in reality is developed within the same 
project and by the same developers, hence, should be counted as own code. As an 
example one could refer to Scala libraries.

Software projects always leverage some functionality
from standard libraries of the programming language, so that one should consider also the size of the the baseline of programming language libraries $\ell_{std}$.

\section{Technical Leverage}
\label{sec:tech:debt}
To capture the effect of software dependencies on the dependent projects, we introduce the 
notion of \textbf{technical leverage}:
\begin{definition}
The \emph{technical leverage} $\lambda$ of a library is the ratio between the 
size of code imported from third-party libraries besides the baseline of programming language libraries $\ell_{std}$
and the own code size of the library: 
\begin{equation}
\lambda = \frac{\ell_{dir} + \ell_{trans} + \ell_{std}}{\ell_{own}}
\end{equation}
\end{definition}
In 
our empirical analysis, the programming language/platform is the same over all 
libraries (Java and Maven) so $\ell_{std}=const$. 
If one wanted to compare libraries across different ecosystems (e.g. 
Python vs C libraries) the difference can be 
significant. Further, standard libraries are typically more mature than a 
third party library and splitting leverage by type might be needed for a more 
fine grained analysis. This is also done in finance where one distinguishes
between  different type of debts.

Similarly to Allman~\cite{allman2012managing}, who drew parallels between 
technical and monetary debts, we illustrate the similarities between financial and 
technical leverages in Table~\ref{tab:tech:fin:leverage}.
\begin{table*}[t]
\caption{Financial leverage vs Technical leverage}
\label{tab:tech:fin:leverage}
\begin{tabularx}{\textwidth}{p{0.4\textwidth} p{0.55\textwidth}}
\hline
Financial leverage & Technical leverage \\ \hline
\rowcolor{Gray}
Financial leverage is used to undertake some investment or
project with the help of borrowed money (debt) & Software developers reuse 
already existing functionality from dependencies to focus only on new features 
in their projects \\

Financial leverage decreases the corporate income tax liability and increases 
its after-tax operating earning~\cite{kraus1973state} & Using dependencies reduce the 
time (and thus,  cost) to develop new 
projects~\cite{baldassarre2005industrial,morisio2002quality} and sometimes 
increase project performance (e.g., \textit{numpy} or \textit{pandas} in Python) \\

\rowcolor{Gray}
The debt implies an interest rate~\cite{kraus1973state} and must be eventually paid \emph{or refinanced} (an observation 
not present in~\cite{allman2012managing}) &  Dependencies have to be monitored and 
updated (the concept similar to refinancing one's monetary debt) unless one introduces security vulnerabilities~\cite{pashchenko2018vulnerable, kula2017ese} \\

Financial leverage multiplies losses as well, which might lead to a 
crisis~\cite{greenlaw2008leveraged} & If the amount of work required for the 
dependent project to update its dependencies become too high, the developers 
might decide to stop updating dependencies and experience serious technical 
difficulties~\cite{pashchenko2020qualitative} \\ \hline
\end{tabularx}
\end{table*}

\begin{figure*}[t]
    \centering
    \includegraphics[width=0.4\columnwidth]{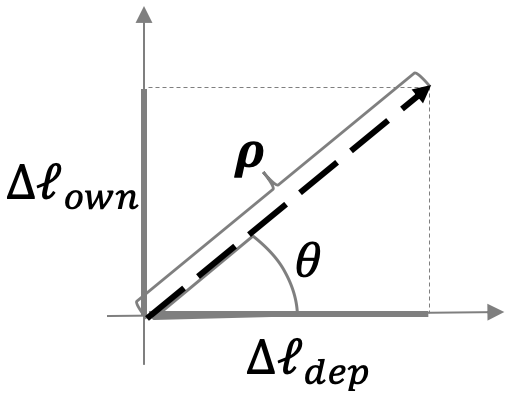}~%
    \includegraphics[width=0.6\textwidth]{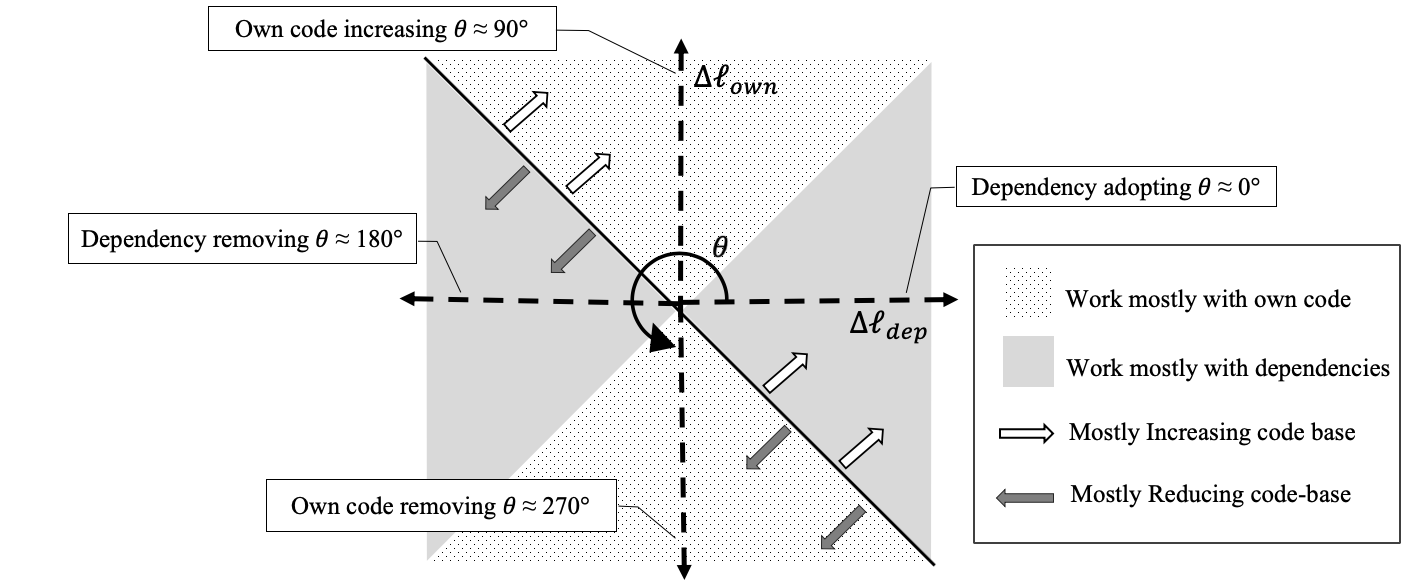}
    \caption{Library change types depending on the angle of a change velocity vector}
    \label{fig:change:types}
    \vspace{-10pt}
\end{figure*}

Additionally, the project maintenance routine requires developers to assess 
whether they have to update dependencies of their project, i.e., to evaluate the 
difference between different versions of the same library. To facilitate this 
process and help developers to have more meaningful comparison 
of changes occurred between two library versions, we propose to use \textbf{change 
velocity vectors} (Figure~\ref{fig:change:types}).

\begin{definition}Change Velocity Vector 
$\langle\Delta\ell_{dep},\Delta\ell_{own}\rangle$ characterizes how a library 
changes between releases $r_0$ and $r_1$ with respect to the size of its
dependencies and its own size:
\begin{equation}
    \langle\Delta\ell_{dep},\Delta\ell_{own}\rangle=\langle\ell_{dep}(r_1)-\ell_{dep}(r_0),\ell_{own}(r_1)-\ell_{own}(r_0)\rangle
\end{equation}
\end{definition}

In particular, the library development behavior can be qualitatively described using polar 
coordinates of the change velocity vector $\theta$ and $\rho$.
We use them to introduce the notions of change direction and change distance.
\begin{definition}
The change distance $\rho$ characterizes the amount of change in code size between two consecutive library version.\begin{equation}
\rho = \sqrt{\Delta\ell_{own}^2+\Delta\ell_{dep}^2}
\end{equation}
\end{definition}

Albeit still measured in KLoC, this is obviously \emph{different} than the pure change in the code base captured by $\Delta\ell_{own}$. Table~\ref{table:stats} and Table~\ref{table:stats:metrics}, later in Section~\ref{sec:data}, show that $\rho$ is on average larger than the library total size $\ell_{own}$.

\begin{definition}
The change direction $\theta$
characterizes the type of evolution of a library 
between two consecutive versions.\footnote{Formally, when $\rho=0$ we assume that $
\theta=0$. Experimentally, $\rho$ is never zero as some changes are always present when one releases a new version.}
\begin{equation}
    \theta = \arccos\left(\frac{\Delta\ell_{dep}}{\rho}\right) * \left\{\begin{array}{ll}+1 & \Delta\ell_{own}>0\\
    -1 & \mbox{otherwise}
    \end{array}\right.
\end{equation}
\end{definition}

Considering different values of
$\theta$, we identify four main 
directions of a library evolution (Figure~\ref{fig:change:types}):
\begin{compactitem}
    \item Dependency adopting ($\theta \approx 0^o$) - software developers 
increase the size of library dependencies, while not changing its own size: 
$\Delta \ell_{dep} > 0, \Delta \ell_{own} \rightarrow 0$
    \item Own size increasing ($\theta \approx 90^o$) - developers do not change 
the dependency size, while increasing its own size: $\Delta \ell_{dep} 
\rightarrow 0, \Delta \ell_{own} > 0$
    \item Dependency removing ($\theta \approx 180^o$) - software developers 
decrease the dependency size, while not changing its own size: $\Delta 
\ell_{dep} < 0, \Delta \ell_{own} \rightarrow 0$
    \item Own code removing ($\theta \approx 270^o$) - developers do not change 
the dependency size, while decreasing own size of their libraries: $\Delta \ell_{dep} 
\rightarrow 0, \Delta \ell_{own} < 0$
\end{compactitem}

Combination of these library evolution directions can describe every change of a 
library. For example, if both own and dependency sizes increase between two 
library releases ($\theta \in (0^o;90^o)$), one may say that its
developers both adopt new dependencies and perform self-development of this 
library. Hence, we qualitatively classify developers' actions as follows:
\begin{compactitem}
\item $\theta \in (315^o;45^o] \cup (135^o;225^o]$ -- developers mostly operate 
with dependencies: $\Delta\ell_{dep} > \Delta\ell_{own}$ (grey area in 
Figure~\ref{fig:change:types})
\item $\theta \in (45^o;135^o] \cup (225^o;315^o]$ -- developers mostly change own 
code: $\Delta\ell_{own} > \Delta\ell_{dep}$ (dotted area in 
Figure~\ref{fig:change:types})
\end{compactitem}

Moreover, the change velocity angle $\theta$ indicates whether 
developers increase or decrease the total size of their libraries:
\begin{compactitem}
\item $\theta \in (315^o;135^o]$ -- developers increase total size $\ell_{total}\uparrow$
\item $\theta \in (135^o;315^o]$ -- developers decrease total size $\ell_{total}\downarrow$
\end{compactitem}

\section{Data selection}
\label{sec:data}


For the study of the \OSS\ ecosystem, we have selected the Maven ecosystem.
We use the information directly available from the dependency management system. 
The \OSS libraries distributed via Apache Maven are published on the Maven 
Central Software Repository\footnote{\url{https://repo.maven.apache.org/maven2/}}, 
that keeps all the publicly released versions of its libraries, i.e., their 
packages (for example, jar), project object model files (pom-files), and, often, 
some extra information, such as source code of a library or its documentation 
(JavaDoc). Maven also provides a Dependency plug-in, that allows us to retrieve 
a list of dependencies of a particular library instance. 

In this study, we use only \emph{direct dependencies}, since
developers have a habit of reacting to
the issues connected with the own code of their libraries or their direct 
dependencies~\cite{derr2017keep, pashchenko2020qualitative}.  Moreover, 
including the analysis of transitive dependencies increases the chance of 
introducing additional biases as transitive vulnerabilities are known to be 
overcounted~\cite{pashchenko2018vulnerable, pashchenko2020vuln4real}.
We do \textit{not} preclude the analysis of transitive 
dependencies as they are known to introduce security 
vulnerabilities to some extent~\cite{blackduck, hejderup2015dependencies,  kula2017ese}.  
Proper understanding of their effects 
requires networks and contagion analysis~\cite{glasserman2016contagion}, 
an interesting challenge for future work 
(Section~\ref{sec:ecosystems:conclusions}).

To identify the relevant `main libraries' and compute their leverage 
one needs a reference point (anchor) selected from the outside of the analysed 
ecosystem.  Indeed, just using the number of usages of within Maven itself would 
have been severely biased as it would not correspond to the popularity of the 
software in the world (which is what makes the study interesting), but only to 
the internal use.  Hence, selecting an anchor from within an ecosystem, we would 
have some important libraries (e.g., Apache Tomcat) underrepresented in the 
library sample, while several service libraries (which nobody really uses) would 
have been disproportionately selected. 


Hence, we follow \cite{Anonymous} and 
started from the top 200 \OSS\ Maven-based libraries used 
by a large software manufacturer across over 500 Java projects 
(actual customers products or production-level software 
developed by the company for internal use).
The resulting set corresponds to 
10905 library instances when considering all versions and includes widely used 
libraries such as \code{org.slf4j:slf4j-api} and 
\code{org.apache.httpcomponents:httpclient}.
\begin{table}[t]
\centering
\caption{Descriptive statistics of the library sample}
\label{table:stats}
\vspace{-5pt}

\longcaption{\columnwidth}{We consider 8464 distinct library versions (GAVs - 
Group, Artifact, Version coordinate in Maven Terminology) starting from the 200 
most popular \OSS Java libraries (GA) used by a multinational software 
development company in its customers or internal production-level software and removing the library versions that 
have no source code in Maven Central.}
\scriptsize
\begin{tabular}{lrrrrrrr}
\hline
& mean  & st.dev & min & $Q25\%$  & median  & $Q75\%$ & max      \\ 
\hline
\#lib versions            
& 55 & 49   & 1  & 15 & 35   & 87 & 248   \\ 
\#direct deps 
& 4  & 7    & 0 & 0 & 2      & 6 & 51      \\ 
$\ell_{own}$ (KLoC) 
& 37  & 56   & 2 & 5 & 15 & 42 & 350  \\ 
$\ell_{dep}$ (KLoC)
& 591  & 764  & 0 & 69  & 302    & 828  & 4489      \\
rel\_interval (days) & 41 & 94 & 0 & 1 & 22 & 47 & 2235  \\ \hline
\end{tabular}
\vspace{-10pt}
\end{table}


Algorithm~\ref{alg:own:size} is used to identify the own size of a library 
instance. To calculate the size of library dependencies, we recursively apply 
the Algorithm~\ref{alg:own:size} to each dependency and then sum the resulting 
number of lines of code (LoCs).

\begin{algorithm}[t]
\vspace{\baselineskip}
\footnotesize
\SetKwInOut{Input}{input}\SetKwInOut{Output}{output}

\Input{A folder $dir$ with the source code of a library}
\Output{The number of lines of code in a library $num\_locs$}

$file\_list \leftarrow getAllFileNames(dir)$ \tcp{Get the list of all file names in the folder $dir$}

$num\_locs \leftarrow 0$\;
\For{$file|file \in file\_list$}{
    \tcp{Counting the number of lines in a file}
    $lines \leftarrow readAllLines(file)$ \tcp{Load content of a file}
    \If{$isCodeFile(file)$}{
        \tcp{Including only code containing files}
        \For{$line \in lines$}{
        \tcp{Counting only lines that are not empty and are not comments}
            \If{$line <> \emptyset$ and $isNotComment(line)$}{
               $num\_locs \leftarrow num\_locs + 1$\;
               }
        }
    }
}
\caption{Extract own size of a library version}
\label{alg:own:size}
\end{algorithm}


For some library instances (or their dependencies) there was no source code 
available, so we removed them from our analysis. We have also removed 555 library 
versions with own size $< 100$ lines of code as these library versions do not 
carry actual functionality but only serve as APIs or documentations for other 
libraries.  The final list comprises 
8464 library instances. Table~\ref{table:stats} presents the
descriptive statistics of the selected library sample.

Developers of some libraries maintain several versions of the library at 
the same time\footnote{In our sample, 570 libraries supported several 
versions simultaneously for a total of 15814 chains.}. For example, the 
developers of Apache
Tomcat\footnote{\url{http://tomcat.apache.org/}} project supported four versions 
(7.0.x, 8.0.x, 8.5.x, and 9.0.x) of 
\code{org.apache.tomcat:tomcat-catalina} library for the last three years 
(starting 
from March,  2016).  These parallel versions might introduce 
errors into the analysis results if date ordering of library releases is 
used.  Ordering versions by date we obtain: 
8.5.30-\textgreater{}9.0.7-\textgreater{}7.0.86-\textgreater{}8.0.51-\textgreater{}9.0.8-\textgreater{}8.5.31 
because versions are released in groups (typically a simultaneous fix of bugs): 
8.5.30 and 9.0.7 were released within 17 minutes. Hence, before the analysis we 
have distinguished release chains according to Algorithm~\ref{alg:dep:split}.
\begin{algorithm}[t]
\footnotesize
\SetKwInOut{Input}{input}\SetKwInOut{Output}{output}

\Input{A set of library names $libraries$}
\Output{A set of lists of consecutive releases $releases$}

$releases \leftarrow []$\;
\For {$library \in libraries$} {
	$cur\_lib = library.getGA()$ \tcp{Use groupId:artifactId as identificator for 
the current library}
	$releases[cur\_lib] \leftarrow []$\;
	$branches \leftarrow []$\tcp{Prepare a list for storing library branches}
	\For {$i \in range(0, len(library))$}{
		$lib\_version \leftarrow library[i]$ \tcp{get i-th library instance of a 
library}
		\If {$releases[cur\_lib] == \emptyset$}{
			$releases[cur\_lib] \leftarrow [lib\_version]$\;
		}
		$lib\_v\_id = cur\_lib + lib\_version[0]$ \tcp{Calculate id of a library 
version}
		\eIf {$lib\_v\_id \in branches$}{
			$releases[lib\_v\_id].append(lib\_version)$\;
		}{
			\eIf {$lib\_version < releases[cur\_lib][-1]$}{
				$branches.append(lib\_v\_id)$\;
				$releases[lib\_v\_id] = [lib\_version]$\;
			}{
				$releases[cur\_lib].append(lib\_version)$\;
			}
		}
	}
}
\caption{Extract consecutive release chains from a library set}
\label{alg:dep:split}
\end{algorithm}

To estimate security risks, we use the presence of a security 
vulnerability that affects the analysed library. 
Here, we consider only 
vulnerabilities that affect the own code of analysed libraries for the quality 
assessment and not the one coming through transitive dependencies. 
In other words, when claiming that a library is vulnerable we will
only count the library own vulnerabilities and the vulnerabilities of the direct
dependencies (which would be own vulnerabilities for each dependency).
At first, this avoids double counting (as the same library may be transitively included several
times).  Second, the presence of vulnerabilities in the direct dependencies
is a knowledge available to a developer. Thus, there is a potentially
deliberate choice of selecting a new vulnerable direct dependency or keeping 
vulnerable dependencies outdated~\cite{pashchenko2020qualitative}.

To identify whether the own code of a library is affected by 
vulnerabilities, we used the Snyk 
database\footnote{\url{https://snyk.io/vuln}} that is constantly updated and for 
August 2020 contains data about more than 3900 vulnerabilities on the Maven based 
libraries. Each entry in the 
database contains the information about a security vulnerability; the library, which 
own code is affected by the vulnerability; and the range of affected 
library versions. Table~\ref{table:stats:vulns} shows
the descriptive statistics of 
vulnerabilities per individual library in the selected library sample.
\begin{table}[t]
\centering
\caption{Descriptive statistics of the vulnerabilities in the sample}\label{table:stats:vulns}
\scriptsize
\begin{tabular}{lrrrrrrr}
\hline
& mean  & st.dev & min & $Q25\%$  & median  & $Q75\%$ & max      \\ 
\hline
\#vulns/own & 2  & 5    & 0    & 0 & 0     &  1 & 43  \\ 
\#vulns/dep & 8  & 10    & 0    & 0 & 3     & 13  & 63  \\ \hline
\end{tabular}
\end{table}

Finally, Table~\ref{table:stats:metrics} shows the descriptive statistics of the proposed 
metrics for the selected library sample.
\begin{table}[t]
\centering
\caption{Descriptive statistics of the proposed metrics}
\label{table:stats:metrics}
\vspace{-5pt}
\longcaption{\columnwidth}{We report the values of the introduced metrics for the  
8464 library instance in our sample for which source code could be extracted and had at least
100 lines of own code (i.e. where not clearly just APIs for other libraries).}
\scriptsize
\begin{tabular}{lr r r r r r r}
\hline
                  & mean  & median & st.dev & min & max    & $Q25\%$   & $Q75\%$   \\ 
\hline
$\lambda_{dir}$ & 2489 & 22 & 21775 & 0 & 373195 & 4 & 74  \\
$\rho$ (KLoC) & 98 & 8 & 284 & 0 & 3480 & 0.71 & 66  \\
$\theta$ (degrees) & 81 & 31 & 91 & -45 & 315 & 0 & 180  \\
 \hline
\end{tabular}
\vspace{-10pt}
\end{table}
In the Section~\ref{sec:sec:rq3}, we will use a leverage of four as a \emph{running example}
 for the impact of direct leverage 
on small and medium libraries which is the reported value for industry average~\cite{blackduck}, it is the value for Q25\% in our sample (Table~\ref{table:stats:metrics}), and corresponds
to the log-mid point between a leverage of 1 (almost 
all libraries are above it) and 16 (over half of the the libraries are above it) as 
visible from Figure~\ref{fig:dep_vs_own}. The equivalent number for 
big libraries would be 12.5\% (log mid point between the median direct leverage of 50\% and the 
bottom line of 1\%).


\section{RQ1: Difference in direct leverage and direction of changes between libraries}
\label{sec:rq1}

Several studies~\cite{humphrey1995discipline, brooks1995mythical, 
aguilar2014size} suggested that a software project might have different 
development practices depending on its size. Hence, we present results separately for \textit{small-medium} libraries 
whose code size does not exceed 100K lines 
of code (KLoCs) and \textit{large} libraries that have more than 100 KLoCs.

\begin{figure*}[t]
    \centering
    \includegraphics[width=\textwidth]{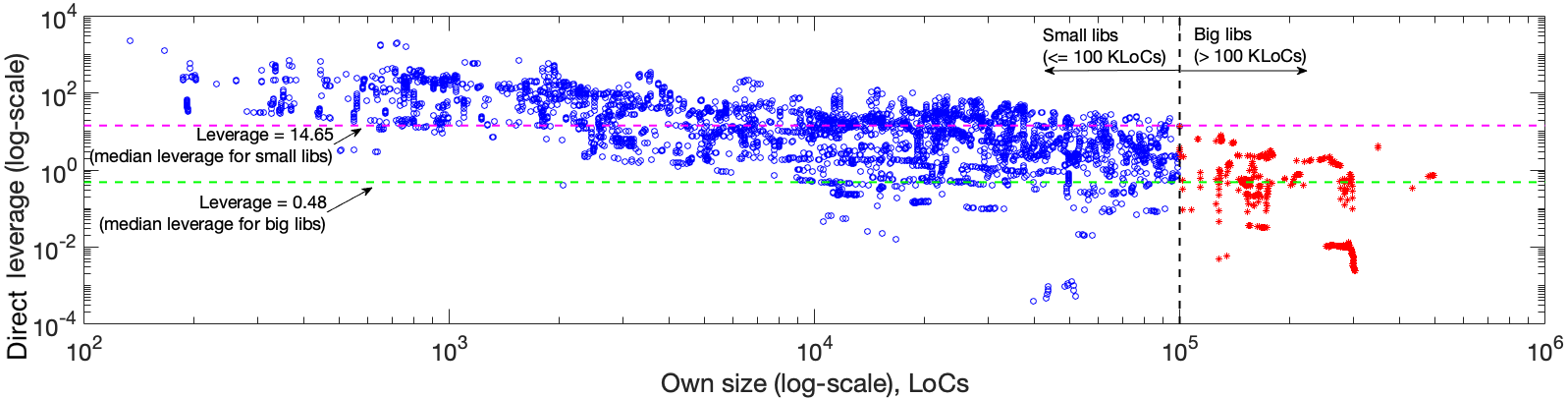}
    
    \longcaption{\textwidth}{Developers of relatively small software
libraries (with own size smaller than 100 KLoCs) almost always ship more code than
their own direct leverage \textgreater{} 1). For the majority, their own code is only
a small fraction of the overall codebase (less than 6\%, corresponding to a median direct leverage of 15). In other words, they ship mostly 
somebody else code. The direct leverage of large libraries (\textgreater{} 100 KLoCs) 
is typically much smaller than the size of their own code and hardly exceed 2, corresponding to at least 33\% of own code.}
    \caption{The direct leverage in comparison to the own size of a library}
    \label{fig:dep_vs_own}
    \vspace{-10pt}
\end{figure*}
\begin{figure*}[ht]
    \centering
    \subfloat[$\theta$ for Small and Medium libraries: The picks at the KDE for the angles of library 
evolution plots suggest that developers of libraries with own code size smaller 
than 100 KLoCs tend to operate with their dependencies: they mostly adopt new 
dependencies and sometimes consolidate them.\label{fig:angles:1}]{
    \includegraphics[width=\textwidth]{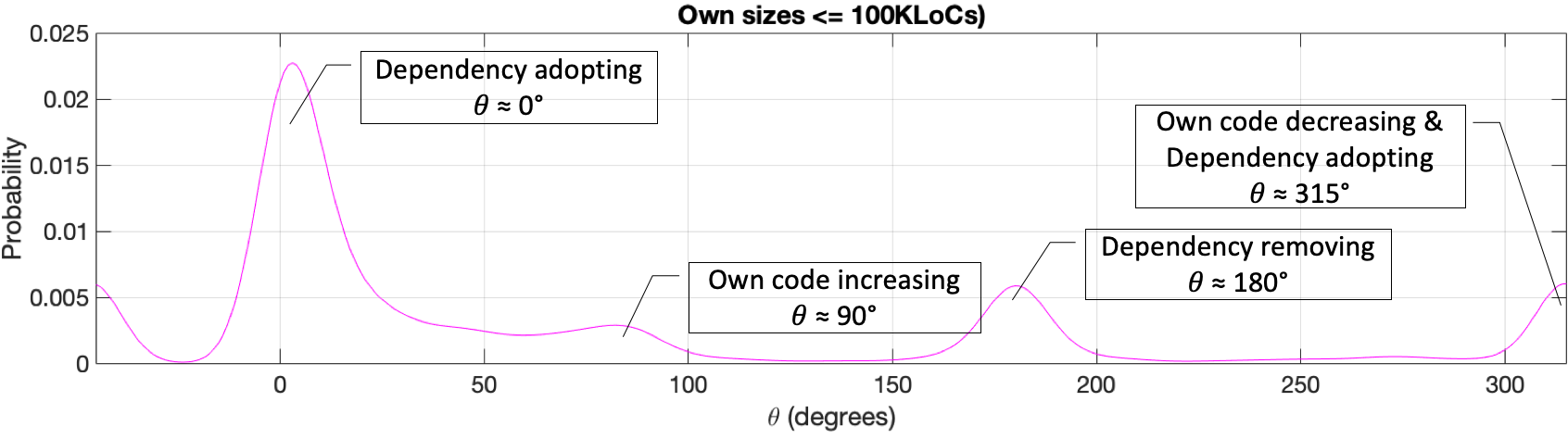}
}


\subfloat[$\theta$ for Large libraries: The KDE of the library evolution vectors for the 
libraries bigger than 100 KLoCs suggest that developers of such libraries tend 
to increase the size of own code while importing some 
functionality from new dependencies (both adopting new dependencies and 
upgrading currently used ones). \label{fig:angles:2}]{
    \includegraphics[width=\textwidth]{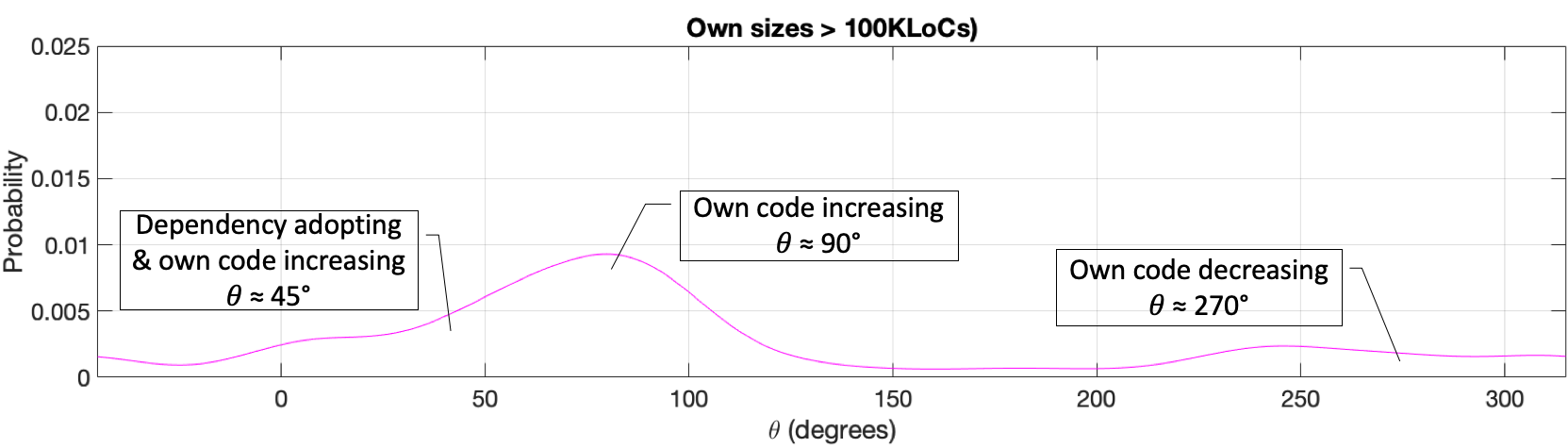}    
}
    \caption{Kernel density estimation plots for angles of library evolution vectors}
    \label{fig:angles}
    \vspace{-10pt}
\end{figure*}


As we observe from Table~\ref{table:stats}, \OSS developers widely adopt 
dependencies to reduce their development effort. This especially applies for the 
small-medium libraries with a code base less than 100 KLoCs
(Figure~\ref{fig:dep_vs_own}): we observe that library instances use a large 
code base of direct dependencies that may 10.000 times exceed their own size and 50\% of 
small-medium libraries rely on 14.65 times bigger code base of their direct dependencies 
($\lambda_{{median}_{\text{small libs}}}=14.65$), which is bigger than the 
direct technical leverage of any library in our sample with own size exceeding 100 
KLoCs. 

The increase of the own size of a library obviously leads to 
a decrease in its leverage: the Pearson 
correlation test~\cite{benesty2009pearson} suggests that there exists linear 
negative correlation between $\log(\lambda_{dir})$ and $\log(\ell_{own})$ ($r=-0.629; 
p\text{-}value \ll 10^{-4}$). From the selected library sample, 50\% of big 
libraries have direct technical leverage less than half of their own size 
($\lambda_{{median}_{\text{big libs}}}=0.48$), including several libraries that 
have 10.000 times smaller size of software dependencies.


Considering the kernel density estimation (KDE) plot for the change velocity 
angle $\theta$ for the libraries in our sample (Figure~\ref{fig:angles}), we 
observe the following:
\begin{itemize}
    \item \textit{small-medium libraries ($\ell_{own} \leq 100 KLoCs$)} 
(Figure~\ref{fig:angles:1}): the developers mostly prefer to adopt third-party 
dependencies (the pick around $\theta \approx 0^o$), rather than focus only 
on increasing the code of their own libraries (the KDE for $\theta \approx 90^o$ 
is less than 0.005). Sometimes developers also reduce 
the size of their dependencies and/or optimize the size of their libraries (the 
KDE for $\theta \approx 180^o$ and for $\theta \approx 315^o$ are higher 
0.005).
    \item \textit{large libraries ($\ell_{own} > 100 KLoCs$)} 
(Figure~\ref{fig:angles:2}): the developers tend to increase the size of their 
libraries (the KDE for $\theta \approx 90^o$ is 0.010), however, the maximum KDE 
value for the large libraries is more than 2 times lower than the maximum KDE for the 
small-medium libraries. Moreover, we observe that the developers also 
sometimes adopt new dependencies but they simultaneously increase own code of 
their libraries (the KDE for $\theta \approx 45^o$ is around 0.005).  Sometimes 
developers of big libraries also optimize their libraries by reducing their own 
code and the code of their dependencies (a pick around $\theta 
\approx 250^o$).
\end{itemize}


\noindent \textbf{Discussion:} Both analysis of direct leverage and change direction
suggest that the developers of small and medium libraries rely 
on functionality of third-party \OSS dependencies. This corresponds to a general 
intuition since the third-party dependencies allow small libraries to grow 
faster. Too many dependencies might become difficult to manage, which 
exposes dependent libraries to bugs and security vulnerabilities 
introduced by library dependencies. Hence, developers of bigger (and 
consequently more mature) libraries tend to decrease the direct leverage of their 
libraries on third-party dependencies: the negative linear correlation between 
$\log(\lambda_{dir})$ and $\log(\ell_{own})$ (Figure~\ref{fig:dep_vs_own}) and a shift 
of development preferences from dependency adopting in small and medium 
libraries to mostly increasing the own code for large libraries 
(Figure~\ref{fig:angles}).

\vspace{0.5\baselineskip}
\fbox{\parbox{0.9\columnwidth}{\textbf{Key Takeaway:} small-medium libraries by 
far ship other people's code, and at each round they add more.}}

\section{RQ2: Does direct leverage impact time interval between library releases?}
\label{sec:sec:rq2}

Big software projects typically involve developers working in 
parallel~\cite{aguilar2014size} to increase the speed of development. To
capture this phenomena, we will use $\log(rel\_interval + 1)$ as 
the dependent variable for the regression. We add `$+1$' to a time interval
between library releases to correct for the approximate granularity in the measurement
as in some cases releases happen on the same day (although at different times).


A large direct leverage means that a library mostly relies on the work of 
other developers. So we expect that direct leverage and library release 
time interval have a proportional relation rather than a linear one.  Since we define 
$\lambda$ as a fraction of dependency and own size of a library, we could use 
$\lambda_{dir}$ directly in the regression.  However,  this would mean that moving 
direct leverage by a factor of 10 would yield an exponential jump in a 
time interval between library releases.
However, we observe that 
direct leverage spans eight orders of magnitude (Fig~\ref{fig:dep_vs_own}) while the 
leverage coefficient in Table~\ref{tbl:rq1:2} suggests that a change in 
magnitude in direct leverage (aka, `effect of scale') does not determine a proportional change in the time interval 
between library releases, only a small linear change.  Therefore, we use 
$\log(\lambda_{dir})$ for the regression.


We expect that the change distance in own and dependency sizes will mostly 
require an increase of the time to test the added functionality into a new 
library version,  i.e.  a `returns to scale' effect.  Hence,  we use $\log(\rho$) for the 
regression.  Notice that $\rho > 0$ as a new release 
always implies some changes.


Change direction indicates (i) how developers change the total size of their 
libraries (increase or decrease) and (ii) whether they mostly change a library's own 
code or its dependencies. To capture the effect of change in 
the total code size, we consider the following:
\begin{compactitem}
\item $\theta = 135^o$ and $\theta = 315^o$ correspond to $0$ change in 
$\ell_{total}$, while $\theta = 45^o$ and $\theta = 225^o$ indicates max 
change in $\ell_{total}$
\item $\ell_{total}$ increases when $\theta \in (315^o; 45^o) \cup (135^o; 225^o)$ 
and decreases when $\theta \in (225^o; 315^o) \cup (45^o; 135^o)$.
\end{compactitem}

To consider these effects into the regression, we introduce the $\cos(\theta - 
45^o)$ transformation of the change direction metric.

If developers change own code of their libraries they have to both 
develop and test their projects. On the other hand, the change of library dependencies 
in most cases require only testing of dependent projects. To capture these effects into 
the regression, we consider the values of $\theta$ that indicate whether 
developers work with dependencies or own code of their libraries and introduce 
the $\sin(\theta)$ transformation of 
the change direction metric\footnote{One might argue that $\cos(\theta)$ and 
$\sin(\theta)$ correlate, and therefore, could not be used as variables into a 
linear regression. However, Eubak and Speckman~\cite{eubank1990curve} proved 
that $\cos(\theta)$ and $\sin(\theta)$ can be used simultaneously in a linear 
regression model and $\cos(\theta+d)$ can be transformed in a linear 
combination of $\cos(\theta)$ and $\sin(\theta)$.}.

Finally,  we consider the time interval between the previous 
release and its preceding one with the corresponding transformations as for the 
current time interval between library releases 
into the regression to capture the impact of project release practices (e.g., 
bi-/weekly/daily releases).

We use the multivariate linear regression 
model~\cite{anderson1962introduction} to check the correlation between the 
proposed metrics and the time 
interval between library releases.  The previous discussion 
clarifies how the transformations above may help towards a
 linear regression model as the direct relation between the proposed
metrics and the estimated parameters might not be necessarily linear (e.g., 
consider the example of Figure~\ref{fig:dep_vs_own}).

The resulted linear regression model has the following form:
\begin{eqnarray}
\log(rel\_interval + 1) \sim 1 + \log(rel\_interval\_prev + 1) + & \hspace*{-5ex}\nonumber \\
&\hspace*{-50ex}+ \log(\lambda_{dir}) + \log(\rho) + \cos(\theta - 45^o) + \sin(\theta) 
\label{eq:regression}
\end{eqnarray}

Table~\ref{tbl:rq1:2} shows the estimates, standard errors, t-statistics, and 
p-values for both small-medium and large libraries. We observe that for the  
libraries with own code smaller than 100 KLoCs all metrics have significant 
impact on the time interval
between library releases ($p\text{-}value < 0.05$), 
Direct leverage,  the change in 
total code,  and previous release have positive correlation with the
release time interval
of a library,  while the change distance and change in own code correlate 
negatively with such interval.
For large libraries,  the change distance and time interval between the previous
and its preceding releases have significant impact on time interval between
library releases ($p\text{-}value < 0.05$).
\begin{table*}[th]
\centering
\caption{Linear model fit to check the correlation between $\theta, \rho, \lambda_{dir}$ and
release time interval}
\label{tbl:rq1:2}
\vspace{-10pt}
\longcaption{\textwidth}{ These are the result of the regressions 
$\log(rel\_interval + 1) \sim 1 + \log(\lambda_{dir}) + \log(\rho) + \cos(\theta - 45^o) + 
\sin(\theta)+\log(rel\_interval\_prev + 1)$.  For small and medium libraries with own\_size
smaller than 100 KLoCs there is root mean squared error = 1.68; $R^2=0.038$ and 
$\overline{R}^2=0.036$.  For large libraries with own\_size 
greater than 100 KLoCs the root mean squared error = 1.53, 
 $R^2=0.058$,  and $\overline{R}^2=0.044$. 
 }
\begin{tabular}{llrlr@{\hspace{4ex}}lrlrl}
\hline
release time interval && \multicolumn{4}{c}{Small-Medium Libraries} & \multicolumn{4}{c}{Large Libraries $>$100KLoC} 
\\
$\log(rel\_interval + 1)$ &  coefficients & estimate & std.err. & tStat & p-value            & estimate  & std.err. & tStat & p-value
\\\hline           
intercept & 1 & 2,847 & 0,1 & 28,413 & 0 & 3,849 & 0,389 & 9,886 & 0 \\
direct leverage & $\log(\lambda_{dir})$ & 0,059 & 0,012 & 4,849 & 0 & -0,002 & 0,041 & -0,041 & 0,968 \\
change distance & $\log(\rho)$ & -0,072 & 0,01 & -7,022 & 0 & -0,147 & 0,036 & -4,091 & 0 \\
change in total code & $\cos(\theta - 45^o)$  & 0,492 & 0,07 & 7,011 & 0 & -0,182 & 0,219 & -0,829 & 0,408 \\
change in own code & $\sin(\theta)$ & -0,355 & 0,075 & -4,709 & 0 & 0,237 & 0,244 & 0,97 & 0,333 \\
previous release interval & $\log(rel\_interval\_prev + 1)$ & 0,083 & 0,016 & 5,033 & 0 & -0,131 & 0,055 & -2,36 & 0,019 \\ \hline
\end{tabular}
\vspace{-10pt}
\end{table*}

\textbf{Discussion.} The increase of change distance 
corresponds to a slight decrease in the 
time interval between
library releases.  Such change is minor in 
quantity: when change distance doubles, such interval decreases only by 7\%. Since the size of $\rho$ is mostly determined by 
dependencies,  this observation shows that adopting dependencies 
speed up the evolution of a library in spite of large (implicit) changes to the 
code base. 
 
What makes the difference is the type of changes.  For small libraries,  changing 
$\ell_{total} (\cos(\theta - 45^o)$
in the regression) lengthen the 
time interval between library releases. 
Most likely,  we observe 
such an effect, since the change in total code base means the major 
functionality changes (e.g., addition of new functionality). In contrast, when 
developers focus on own code of their libraries,  
such interval reduces by 
35\% -- the developers are likely fixing bugs and vulnerabilities in their 
libraries, which might be of a higher priority due to the necessity to ship the 
fixed version to the library users. A qualitative study might yield more insights
into such changes.

\subsection{Does direct technical leverage pay off?} To understand the 
practical implication of direct technical leverage, we analyze further the impact of the
coefficient of direct technical leverage in (\ref{eq:regression}). 

Since the $\log(\lambda_{dir})$ coefficient is positive, 
having a large direct leverage increases the 
time interval between releases of a small-medium library:
having too many dependencies might 
require additional time for the library developers to manage them. 
However,  the value of the coefficient is small, and therefore,
a more precise estimation is needed. 

Let the $rel\_interval'$ and the 
$rel\_interval$ be two release time intervals of libraries that are separated by
a leverage factor of $\Lambda$,  i.e.  $\lambda'=\Lambda\cdot\lambda$. From the 
regression we can reconstruct the following (approximate) relation:
\begin{eqnarray*}
\log(rel\_interval'+1)-\log(rel\_interval+1) =& \\
=\beta_\lambda (\log(\Lambda\cdot\lambda)-\log(\lambda))
\end{eqnarray*}
 which can be simplified to

\begin{equation}
\frac{rel\_interval'+1}{rel\_interval+1}= \Lambda^{\beta_\lambda}
\end{equation}

Considering the estimates in Table~\ref{tbl:rq1:2},  for a library with 
own code $<100KLoCs$ and direct technical leverage of four (the size of its direct dependencies 
is four times bigger than its own size), the delay in the 
time interval between library releases in days will be less
than 4\%. So, leveraging on third-party libraries pays off: you add 4x more 
code at the price of a small delay in your 
release time interval (around two days on average).
Even if you have many dependencies (e.g. , $\Lambda=16$, slightly above the median
for small libraries, and your own code is essentially 6\% of the total code size), 
the time interval between library
releases only increases by 17\% (around a week
considering the average time interval in Table~\ref{table:stats}).

\vspace{0.5\baselineskip}
\fbox{\parbox{0.9\columnwidth}{\textbf{Key Takeaway:} Direct leverage pays off.  Shipping an overall project four times larger than your own code base will only take a couple of extra days (on average).}}

\section{RQ3: Does direct leverage increase security risk?}
\label{sec:sec:rq3}

%
%

In medicine, the effect of a parameter on a rare disease is described by the 
\textit{odds ratio} $OR$ of a disease~\cite{szumilas2010explaining}.  We 
use the $OR$ to have a first understanding of the impact of leverage on the 
security risk\footnote{As we only know \emph{reported} vulnerabilities for the 
dependencies in our sample (as vulnerabilities might be present but might not have been  found yet), we cannot use the \textit{risk ratio}. However,  since 
vulnerabilities are rare, the OR approximates the risk 
ratio~\cite{allodi2014comparing}.} of using a library, by mapping (i) the fact 
that at least one vulnerability affects a software library onto `Disease' and 
(ii) the condition that the direct technical leverage exceeds our running example value 
for leverage (four for small libraries) onto the state of `Exposure':
\begin{equation}
OR = 
\frac{%
\frac{\mid High Leverage Libs \text{ }\cap \text{ }Vuln Libs\mid}{\mid High Leverage Libs \text{ }\cap \text{ }\neg Vuln Libs\mid}%
}{%
\frac{\mid Low Leverage Libs \text{ }\cap \text{ }Vuln Libs\mid}{\mid Low Leverage Libs \text{ }\cap \text{ }\neg Vuln Libs\mid}%
}
\end{equation}

Table~\ref{tab:odds:ratio} shows the contingency table where we use our running example
 $\lambda = 4$ 
 for small and medium libraries (own code $\leq 100 KLoCs$ - median 
direct leverage around 16) and $\lambda_{dir} = 12.5\%$ for large 
libraries (own code $>100KLoCs$ - median direct leverage around 50\%). 
The corresponding library 
groups have the following odds ratio: $OR_{\text{small libs}} = 1.6$, confidence interval
$[1.3\text{ }2.0]$. For big libraries the 
$OR_{\text{big libs}} = 0.43$ with confidence interval $[0.22\text{ }0.84]$.
Fisher Exact test~\cite{fisher1935logic} for both small-medium and big libraries 
rejects $h_0$ ($p\text{-}value_{\text{small libs}} \ll 0.05$, 
$p\text{-}value_{\text{big libs}} = 0.013$).
\begin{table}
\centering
\caption{Contingency table for vulnerable/leveraged libraries}
\label{tab:odds:ratio}
\longcaption{\columnwidth}{For small and medium libraries, the larger the 
direct leverage the larger the risk of being vulnerable ($OR_{\text{small libs}} = 
1.61$). For large libraries (own code $>100KLoCs$), the risk is inverted, the 
more own code the more likely to be vulnerable ($OR_{\text{small libs}} = 
0.43$).}
\begin{tabular}{lrrlrr}
\hline
 & \multicolumn{2}{c}{Small libraries} & & \multicolumn{2}{c}{Big libraries} \\
 & vuln & not vuln & & vuln & not vuln \\ \hline
$\lambda_{dir} > 4$ & 716 & 2154 & $\lambda_{dir} > 0.125$ & 194 & 74 \\
$\lambda_{dir} \leq 4$ & 121 & 587 & $\lambda_{dir} \leq 0.125$ & 73 & 12 \\ \hline
\end{tabular}
\end{table}
These results confirm our intuition: the more you increase your total code size, 
the more likely you are to step into a vulnerability. The major difference is 
that for small and medium libraries some of these vulnerabilities are most 
likely in other people's code and therefore out of control of the developer.

Since a vulnerable library has normally more than one vulnerability (see 
Table~\ref{table:stats:vulns}), it is interesting to understand whether 
the \emph{number of vulnerabilities} change with leverage\footnote{From a user's 
perspective the library (GA in Maven's terminology) is the `same', it is  just a 
`different version` (GAV in Maven's terminology).}. Further,
a library may have different level of leverage from version to version 
while the number of vulnerabilities may remain unchanged. These two
facts are due to the phenomenon that some vulnerabilities are in the third party 
code (in dependencies) that requires developers of these dependencies to release 
a fixed version of a dependency. In other words, developers might have added a 
new library (left-most peak in Figure~\ref{fig:angles}) while keeping an old 
version of another dependency. 

\begin{figure*}
\includegraphics[width=\textwidth]{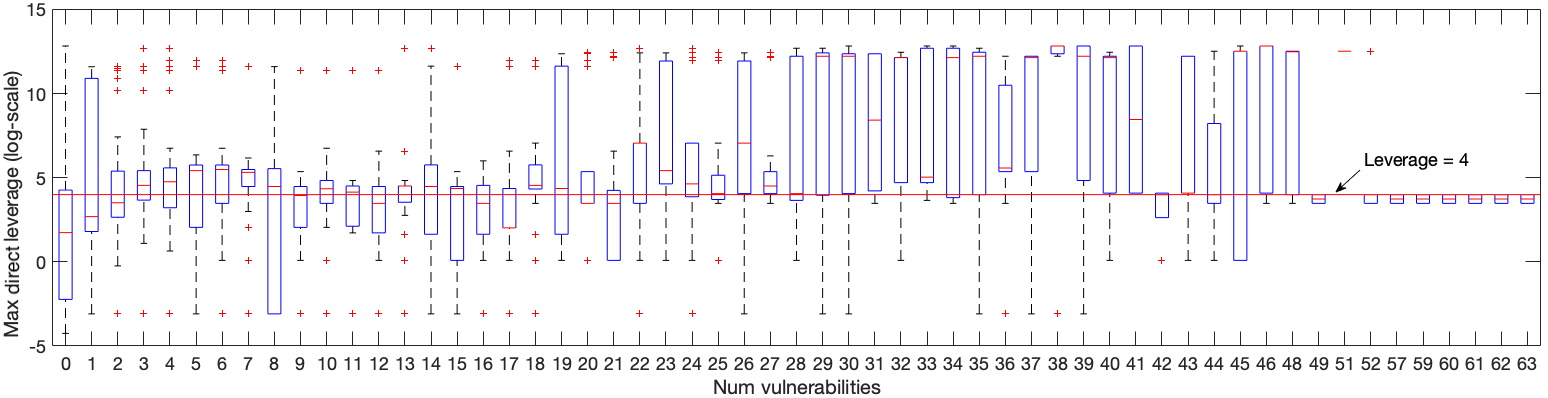}
\longcaption{\textwidth}{Direct leverage equal to 4 allows visual separation 
between the libraries exposed to high number of vulnerabilities vs libraries 
exposed to a small number of security vulnerabilities in our library sample. 
The only excpetion are the handful of libraries at the extreme right. They are libraries with large own code base $\geq 100K$ LoCs which are always affected by security vulnerabilities just because of their size.
}
\caption{Max direct leverage per library vs Number of vulnerabilities in a library version}
\label{fig:max_lev_num_vulns}
\end{figure*}

Figure~\ref{fig:max_lev_num_vulns} shows the distributions of maximum direct leverage
per library in time grouped by the number of vulnerabilities that affect that library (intended as a GA).
Our running example threshold (leverage equal to four)
provides clear visual separation of the exposed and not exposed libraries (i.e., 
libraries with high direct leverage are more likely to be exposed to a security 
vulnerability. 
The exception in Figure~\ref{fig:max_lev_num_vulns} are the libraries with a high number of vulnerabilities (extreme right) but 
a direct leverage lower than 4. They are the big libraries ($\ell_{own}>100K$ LoCs) which are anyway 
exposed to the big number of vulnerabilities due to the large own size. Hence, the direct leverage 
metric has a potential to be used as an indicator for a library to be exposed to 
security vulnerabilities. Further investigations are worth pursuing.

%

\begin{figure*}[t]
\centering
\includegraphics[width=\textwidth]{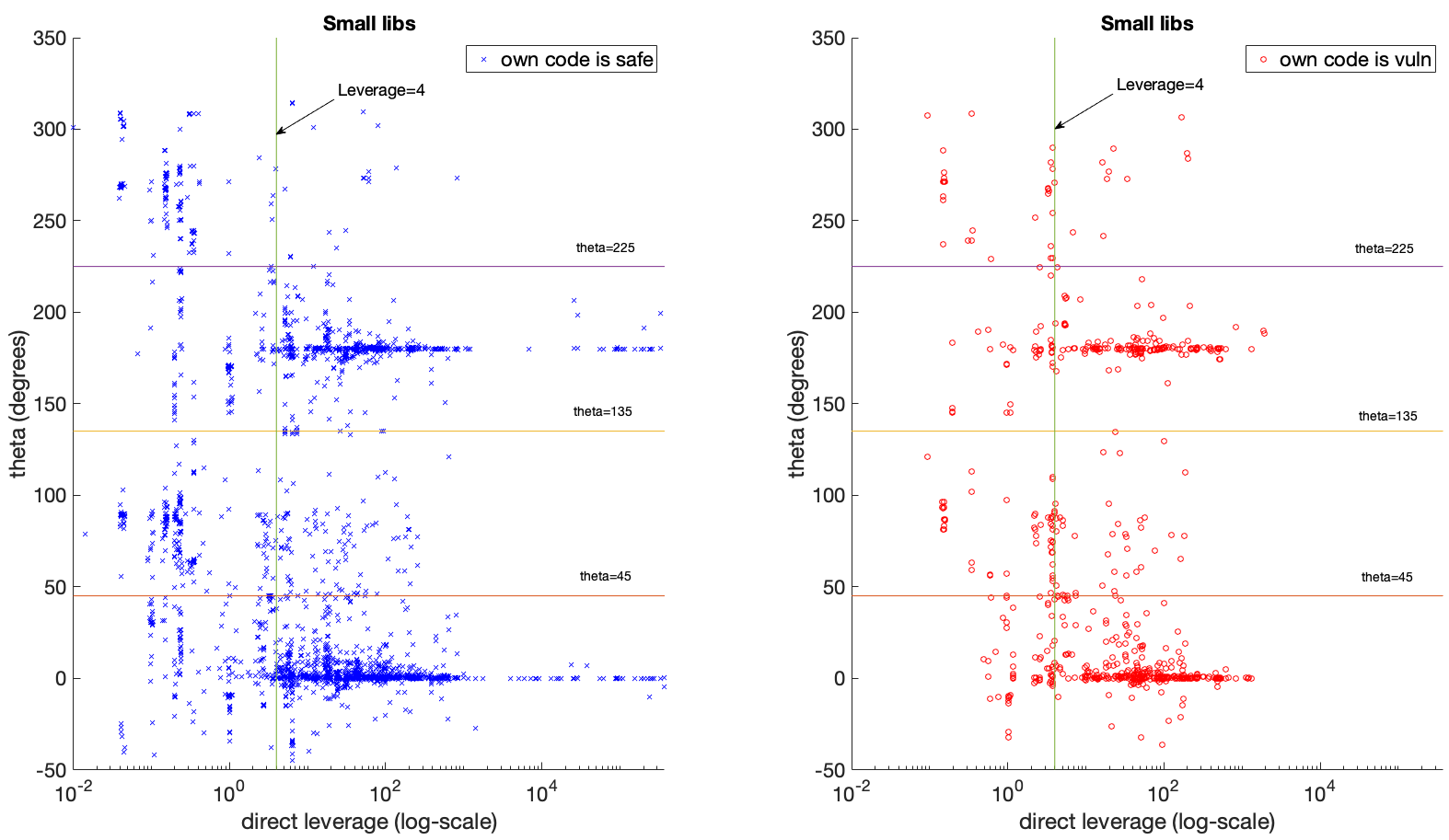}
\vspace{-15pt}
\caption{Direct leverage vs change direction for safe libraries (blue, left) and vulnerable libraries (red, right) with own\_size $\leq 100 KLoCs$}
\label{fig:lev:theta:small}
\vspace{-10pt}
\end{figure*}

Figure~\ref{fig:lev:theta:small} shows the relation between direct leverage and direction of 
library changes ($\theta$). We observe that small-medium libraries (own code $\leq$ 
100 KLoCs) with $\theta \in [-45; 45] \cup [135; 225]$ are more likely to be vulnerable. Such 
libraries either include/remove functionality from software dependencies or 
increase their own code base, and therefore, are likely to be under active 
development. In contract, there are less vulnerable small libraries with $\theta 
\in (225; 315)$. Such libraries decrease the size of their own code, and 
therefore, they are likely to review the already developed functionality instead 
of developing new features (i.e., to be mature).
Visual analysis of the direct leverage--change direction relation plots for libraries with own code 
$>100 KLoCs$ suggests that in case of a big 
library there always exists a chance that its own code is affected by a security 
vulnerability.

\noindent \textbf{Discussion:}
We observe that the fact of being affected by a security vulnerability for 
small-medium libraries correlates with direct leverage and change distance, while for 
big libraries the significant effect comes from change in own and total code. 
Most likely, this happens due to the different strategies followed by library 
developers: the developers of small libraries tend to quickly increase the 
functionality of their libraries, and therefore, adopt new dependencies (which 
increase the total code size of their libraries), while the developers of big 
libraries focus on changes in their own code. 
While changes in own code may increase the attack surface, they normally also 
include eliminating bugs and security vulnerabilities in software 
libraries. In contrast, adding libraries only eliminates security vulnerabilities as a side 
effect if a version (or a whole library) without vulnerabilities is selected.

\vspace{0.5\baselineskip}
\fbox{\parbox{0.9\columnwidth}{\textbf{Key Takeaway:} Direct leverage 
increases the security risk. Shipping four times your code base will increase 
the risk of shipping a vulnerable version by 60\%.}}
\vspace{0.5\baselineskip}

\section{Related works}
\label{sec:rel_works}

Several technical studies~\cite{hejderup2015dependencies, kula2017ese, 
cox2015icse, lauinger2017thou, pashchenko2018vulnerable} showed that \OSS 
dependencies, although being widely used by both commercial and \OSS projects, 
are not often maintained properly: a large share of projects (up to 81\%) have 
outdated dependencies. Several of them (69\%) are not aware that some of those 
dependencies introduce serious bugs and security 
vulnerabilities~\cite{kula2017ese}. As Allman~\cite{allman2012managing} drew 
parallels between technical and monetary debts, one may relate dependencies in 
\OSS to the well-studied financial leverage instruments.
However, we do not find a study that would try to quantitatively assess the 
technical debt introduced by software dependencies.

Manikas and Hansen~\cite{manikas2013software} presented a systematic literature 
review of 90 papers on the studies regarding software ecosystems. Although the 
number of software ecosystem research papers is increasing, the majority of 
studies are report papers. Hence, the authors reported the lack of analytic 
studies of software ecosystems. This statement is supported by another extensive 
literature review of 213 papers on software 
ecosystems~\cite{manikas2016revisiting}. Similar results are found by 
Manikas~\cite{manikas2016supporting} in a more recent literature review of 56 
empirical studies spanning over 55 software ecosystems: there exists a lack of 
deeper investigation of technical and collaborative aspects.

Boucharas et al.~\cite{Boucharas2009FSE} proposed a standards-setting approach 
to software product and software supply network modelling. Although this allows 
developers to anticipate upcoming changes in the software ecosystems, the 
approach aims at development within one company, and therefore, does not suit 
the purpose of modelling \OSS infrastructure.

Bonaccorsi and Rossi~\cite{bonaccorsi2003open} proposed a simple model to 
helps software developers to decide whether to include \OSS components into 
their projects. Their model estimates the value of \OSS libraries based on the possibility
of receiving additional support from the developers of an \OSS community.

Hence, our study fills an important gap in the state of the art by providing 
instruments for evaluation of the impact of technical leverage in the software 
ecosystems.

\section{Threats to Validity}
\label{sec:ecosystems:threats}

The internal validity may be influenced by the fact that we have based the \OSS
library selection for this study on their popularity from within a company. We 
surveyed the usage data of the selected sample from 
MVNRepository.com\footnote{\url{https://mvnrepository.com/}} and the number of 
users from BlackDuck Openhub\footnote{\url{https://www.openhub.net/}}. Since both 
sources showed that libraries in our sample are also popular among the \OSS 
developers, we believe, the internal validity threat of our study is minimal.

The generalization of these results may be exposed to an external validity 
threat since we considered only Maven based libraries. However, since Maven has the largest share of users between the 
developers in the Java ecosystem\footnote{https://zeroturnaround.com/
rebellabs/java-tools-and-technologies-landscape-2016}, our results reflect the 
practice of the majority of Java developers. In this study, we aimed at 
creating awareness regarding the effects of technical leverage within 
software ecosystems, and therefore, proper case control studies are needed to, 
for example, validate the effect of odds ratio. In this respect, our study is 
easy to replicate for other dependency management systems.

\section[Conclusions]{Conclusions and Future Work}
\label{sec:ecosystems:conclusions}

By extending the metaphore from finance started by the notion of 
technical debt~\cite{techdebtmetaphor2012}, we have introduced the new notion of \textbf{technical leverage}
and some associated code metrics (leverage, change distance, and 
change direction) to capture the relative importance and evolution
of one's own code and third party code into a software library. This notion
is particularly important in today's software ecosystem where homegrown
code is only a fraction of the total code base that is shipped to customers (See
e.g.,~\cite{mack2020security,blackduck,gallivan2001striking} and Table~\ref{table:stats:metrics}).

We have applied 
the proposed metrics to 8464 \OSS\ library instances from Java Maven stemming
from an industry relevant sample of the top 200 libraries used by a large multi-national 
corporation for its customers. The results 
show that small-medium libraries have high leverage on third-party \OSS 
dependencies and their developers prefer to adopt new dependencies to speed up 
the development process. Large libraries have relatively small 
leverage and their developers mostly increase own code of their libraries. 
The proposed metrics correlate with
time interval between
library releases and could be used to estimate the risk of a library to be 
affected by a security vulnerability. Libraries 
whose developers perform operations with (e.g., adopt or remove) dependencies 
tend to be affected by vulnerabilities more often in comparison than libraries 
whose developers mostly change libraries' own code. 

We briefly recap the key findings of our paper as follows:
\begin{compactitem}
\item Small and medium libraries (with less than 100 KLoCs of own code) by far ship other people's code, and at each round they typically add more. With a median direct leverage of 14.65, most libraries in our sample include less than 7\% of homegrown code.   
\item  \textbf{{Direct leverage} pays off}.  Shipping four times the size of your own code base (as 75\% of small and medium libraries do) will only take two extra days on average.
\item \textbf{{Direct leverage} increases the security risk}.  Shipping four times the size of your own code base will increase the odds of shipping a vulnerable version by 60\%.
\end{compactitem}
For sake of comparison, ``if all presently unbelted drivers and right front passengers were to use the provided three point lap/shoulder belt, but not otherwise change their behavior, fatalities to this group would decline by (43 ± 3)\%'' \cite{EVANS1986}.

These findings have also interesting implications for novel research directions in empirical
software engineering and software security economics. 

Within empirical software engineering, 
metrics of developer behavior (e.g., unfocussed contribution 
\cite{meneely2009secure}, different  development 
priorities~\cite{peters2012evaluating}, code 
complexity~\cite{zimmermann2007predicting},
large code changes~\cite{jiang2008comparing}, etc.) are often studied to explain code quality. 
Yet, as we have seen, for small and medium
libraries, developers ship overwhelmingly other people's code.
So their \emph{skills and behavior as coders} which may be captured by the logs of
software repositories only contribute in minimal part 
to the quality of the overall code they ship. In contrast, their \emph{decision making behavior}
about the choice of libraries to be used as dependencies in one's own code
has much larger impact. Unfortunately, such behavior is not equally well documented
and captured by traditional software repositories. It might be captured by NLP analysis of the mailing lists 
or discussion blogs of the \OSS\ project (or company internal mechanisms).

 From the perspective of security economics, our empirical data shows that 
 \emph{technical leverage creates a 
 decision dilemma (a moral hazard in the economics terminology).} The benefit of a large leverage
 are reaped by the developers who can ship more code, i.e. more functionalities,
 with a limited delay of time
 interval between releases which
 is associated with greater profitability \cite{august2013influence} and have been
 shown to have a log-linear relation with costs \cite{huijgens2017effort}. Yet,
 the risk of using a vulnerable software are borne by the users of the library which
 might be hit by hackers if they kept an old, vulnerable version that was however perfectly
 functional from their perspective (See the Equifax data breach\footnote{\url{https://blogs.apache.org/foundation/entry/media-alert-the-apache-software}}).
 Since updating one's software is often not a technically feasible solution, 
 as illustrated by a quantitative study on Android libraries in 
\cite{huang2019up} (almost every second library update broke 
the dependent project) and 
qualitatively explained in~\cite{pashchenko2020qualitative}, the presence of 
such dilemma may require to identify
alternative solutions to software updates.
 
We also plan 
to investigate the broader impact of the proposed metrics.  For example, 
leverage may predict a boundary for the amount of dependencies beyond which maintenance and update become 
unwieldy: How many are too many? This might require to correlate leverage with additional metrics such as the 
number of open issues, the effort of developers etc. 
Also, we may expect that library maintainers prefer 
different development strategies (captured by change direction), 
depending on the stage of maturity of a library and it would be 
interesting to determine whether there is such an effect.
Another important direction for future work 
is the study of the impact of transitive 
dependencies on technical leverage albeit this should be done with care to avoid double or triple counting \cite{pashchenko2018vulnerable}. Above all, it will be interesting to further 
investigate the impact of technical leverage on other programming languages
and software repositories.

\section*{More Information}
For the interested readers, we provide an online demo for computing the proposed metrics for the 
software libraries of this study and others at the following URL:
\url{https://techleverage.eu/}

\section*{Acknowledgments}
\label{sec:acknowledgments}
We would like to thank A.Brucker, G. Kuper and P.Tonella for their insightful comments on early drafts 
of this work. The graphical abstract for this paper is an artwork by Anna Formilan \url{http://annaformilan.com}.
This work was partly funded by the European Union under the H2020 
Programme under grant n. 952647 (AssureMOSS).

\bibliographystyle{IEEETran}
\bibliography{IEEEabrv,short-names,biblio}

\begin{thebibliography}{10}
\bibitem{Anonymous}
Anonymous, ``For blind review,'' To be unblinded in the final version.
\end{thebibliography}

\end{document}